\documentclass[bibyear]{aa} 
\usepackage{epsfig}
\usepackage{amsmath}
\usepackage{subfigure}
\usepackage{natbib}
\usepackage{multirow}
\usepackage{color}
\usepackage{ulem}
\usepackage{longtable}
\usepackage{graphicx}
\usepackage{epstopdf}
\usepackage{booktabs}
\usepackage{txfonts}

\newcommand{\jmst}{J.~Mol.~Struct.}   
\newcommand{\jms}{J.~Mol.~Spectrosc.}   

\begin{document}

\title{Detection of deuterated methylcyanoacetylene, CH$_2$DC$_3$N, in TMC-1 \thanks{Based on observations carried out
with the Yebes 40m telescope (projects 19A003, 20A014, and 20D15). The 40m radiotelescope at Yebes Observatory is operated by the Spanish Geographic Institute
(IGN, Ministerio de Transportes, Movilidad y Agenda Urbana).}}

\author{
C.~Cabezas\inst{1},
E.~Roueff\inst{2},
B.~Tercero\inst{3,4},
M.~Ag\'undez\inst{1},
N.~Marcelino\inst{1},
P.~de~Vicente\inst{3}
and
J.~Cernicharo\inst{1}
}

\institute{Grupo de Astrof\'isica Molecular, Instituto de F\'isica Fundamental (IFF-CSIC), C/ Serrano 121, 28006 Madrid, Spain.
\email carlos.cabezas@csic.es; jose.cernicharo@csic.es
\and LERMA, Observatoire de Paris, PSL Research University, CNRS, Sorbonne Universit\'es, 92190 Meudon, France
\and Observatorio Astron\'omico Nacional (IGN), C/ Alfonso XII, 3, 28014, Madrid, Spain.
\and Centro de Desarrollos Tecnol\'ogicos, Observatorio de Yebes (IGN), 19141 Yebes, Guadalajara, Spain.
}

\date{Received; accepted}

\abstract{We report the first detection in space of the single deuterated isotopologue
of methylcyanoacetylene, CH$_2$DC$_3$N. A total of fifteen rotational transitions,
with $J$ = 8-12 and $K_a$ = 0 and 1, were identified for this species in TMC-1 in the
31.0-50.4 GHz range using the Yebes 40m radio telescope. The observed frequencies were
used to  derive for the first time the spectroscopic parameters of this deuterated isotopologue. We
derive a column density of $(8.0\pm 0.4) \times 10^{10}$ cm$^{-2}$. The abundance ratio
between CH$_3$C$_3$N and CH$_2$DC$_3$N is $\sim$22. We also theoretically computed the principal spectroscopic constants of
$^{13}$C isotopologues of CH$_3$C$_3$N and CH$_3$C$_4$H and those of the deuterated isotopologues of CH$_3$C$_4$H for which we could expect a similar degree
of deuteration enhancement. However, we have not detected either CH$_2$DC$_4$H
nor CH$_3$C$_4$D nor any $^{13}$C isotopologue. The different observed deuterium ratios in TMC-1 are reasonably accounted for by a gas phase chemical model where the low temperature conditions favor deuteron transfer through reactions with H$_2$D$^+$.}


\titlerunning{CH$_2$DC$_3$N in TMC-1}
\authorrunning{Cabezas et al.}

\maketitle

\section{Introduction}

Deuterium fractionation is a well-known process in the dense interstellar medium
which can occur both in the gas phase and on the surfaces of dust particles. This
process allows deuterated isotopic species of interstellar molecules to reach
abundances much higher than the D/H elemental abundance ratio ($1.5\times10^{-5}$
\citealt{Linsky2003}). The high efficiency of deuterium fractionation allows deuterated species to achieve abundances as high as 30-40~\% relative to the
parent species, as occurs with HDCS \citep{Marcelino2005} and CH$_2$DOH
\citep{Parise2006}. Hence, deuterated isotopologues of abundant interstellar
molecules make a significant contribution to the spectral richness of
line surveys. This fact makes the astronomical identification of these
isotopologues of utmost importance, not only to gain knowledge on their
molecular formation pathways or how deuterium fractionation works, but
also to assign unidentified features in line surveys.

Very sensitive broadband line surveys of astronomical sources can now be achieved thanks to new technical developments on radiotelescopes. These surveys have boosted up the number of new molecular
identifications in the last years, because weak lines arising from
low-abundance species and from low-dipole moment species can be now
easily detected \citep{Agundez2021a,Cernicharo2021a,Cernicharo2021b,Cernicharo2021c}.
The negative counterpart of this high sensitivity is the huge number of new
lines which populate the survey, including isotopologues and vibrationally
excited states, in warm environments, of well known species. Hence, discovering spectral features of new molecules requires a previous detailed analysis of the spectral contribution of known species.

Methylcyanoacetylene, CH$_3$C$_3$N, also known as cyanopropyne or
methylpropionitrile, has been detected with high abundance in the cold
dark cloud TMC-1 \citep{Broten1984} and more recently
by \citet{Marcelino2021} using a high sensitivity line survey on TMC-1
gathered with the Yebes 40m radio telescope (see, e.g., \citealt{Cernicharo2021d}). Hence, the deuterated
isotopologues of CH$_3$C$_3$N are good candidates to be observed in
this source using the same line survey. In fact, we have already
detected other singly deuterated isotopologues of species such as
CH$_3$CN, CH$_3$CCH, $c$-C$_3$H$_2$, C$_4$H, H$_2$C$_4$, H$_2$CCN,
HC$_3$N and HC$_5$N \citep{Cabezas2021}.

In this Letter we report the identification of spectral lines of the
deuterated species CH$_2$DC$_3$N in TMC-1. Our search for this molecule is
based on the change in the rotational parameters of CH$_3$C$_3$N produced by
the H/D exchange, which have been obtained by ab initio calculations.
The derived deuterium ratios are confronted to an extended
chemical model including the related deuterated compounds.

\section{Observations}

The Q-band observations of TMC-1 ($\alpha_{J2000}=4^{\rm h} 41^{\rm  m}
41.9^{\rm s}$ and $\delta_{J2000}=+25^\circ 41' 27.0''$)
described in this work were performed in several sessions between
November 2019 and April 2021. They were carried out using a set of new receivers,
built within the Nanocosmos project\footnote{\texttt{https://nanocosmos.iff.csic.es/}},
and installed at the Yebes 40m radio telescope.

The Q-band receiver consists of two high electron mobility transitor cold
amplifiers covering the 31.0-50.4 GHz band in horizontal and vertical polarizations.
The receiver temperature varies from 22 K at 32 GHz to 42 K at 50 GHz. The spectrometers formed
by $2\times8\times2.5$ GHz FFTs provide a spectral resolution of 38.15
kHz and cover the whole Q-band in both polarizations. The receivers and
the spectrometers are described before by \citet{Tercero2021}.

Different frequency coverages were observed, 31.08-49.52 GHz and
31.98-50.42 GHz, which permits to check that no spurious ghosts are
produced in the down-conversion chain in which the signal coming from
the receiver is  downconverted to 1-19.5 GHz, and then splits into 8 bands
with a coverage of 2.5 GHz, each of which being analyzed by the FFTs.

The observing procedure used was the frequency switching mode, with a frequency throw of 10\,MHz or 8\,MHz
(see, e.g., \citealt{Cernicharo2021d,Cernicharo2021e,Cernicharo2021f}).

The intensity scale, antenna temperature ($T_A^*$), was calibrated using two absorbers at different temperatures
and the atmospheric transmission model ATM \citep{Cernicharo1985, Pardo2001}.
Calibration uncertainties have been adopted to be 10~\% based on the
observed repeatability of the line intensities between different observing runs. All data have been
analyzed using the GILDAS package\footnote{\texttt{http://www.iram.fr/IRAMFR/GILDAS}}.

\section{Results}

\begin{figure*}
\centering
\includegraphics[angle=0,width=0.90\textwidth]{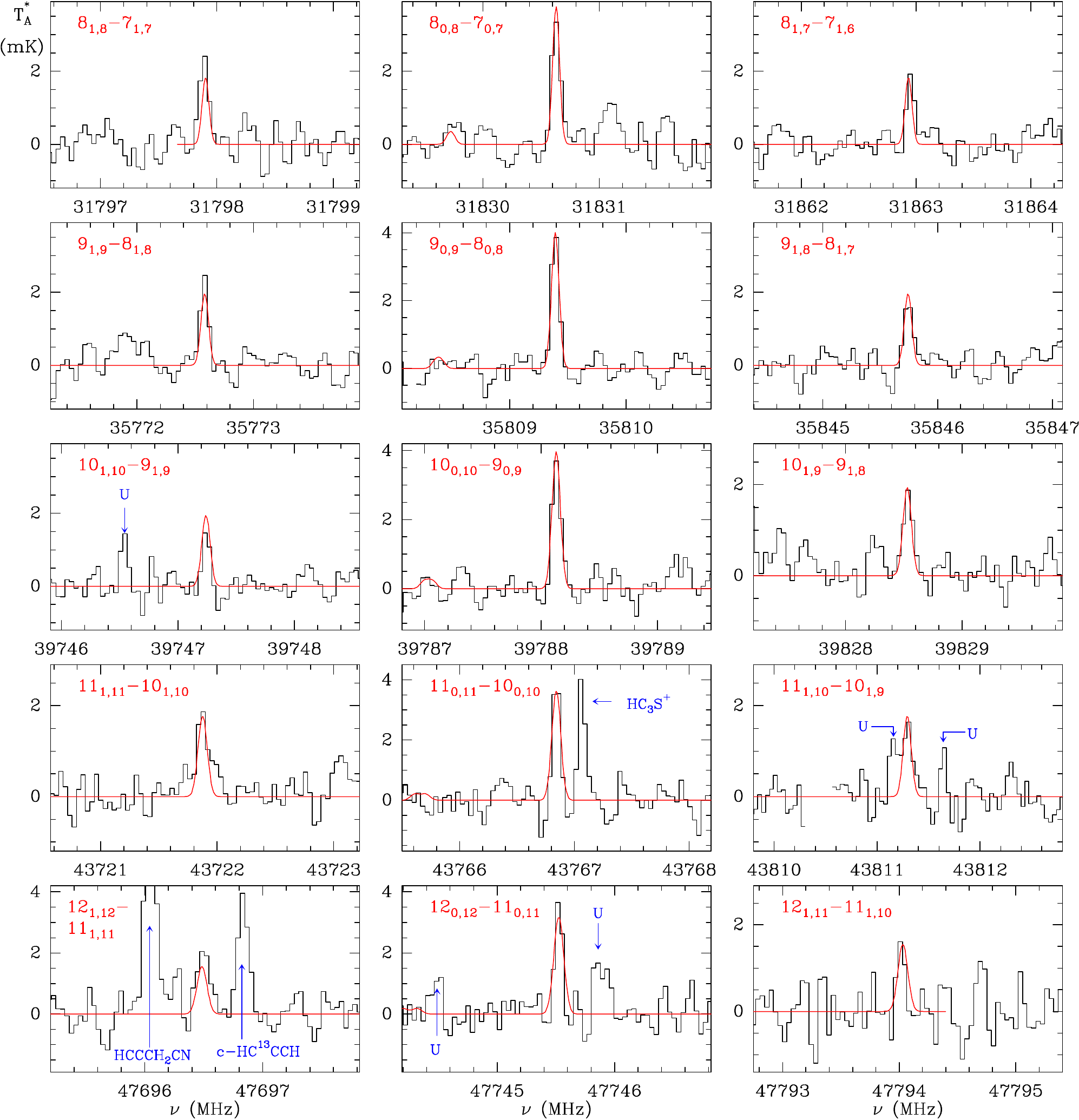}
\caption{Observed lines of CH$_2$DC$_3$N in TMC-1 in the 31.0-50.4 GHz range.
Frequencies and line parameters are given in Table \ref{freq_lines}. Quantum numbers
for the observed transitions are indicated in each panel.
The red line shows the synthetic spectrum computed for a rotational temperature
of 8\,K and a column density of 8$\times$10$^{10}$ cm$^{-2}$ (see text). The additional components seen in the synthetic spectrum close to the $K_a$\,=\,0 components are the $K_a$\,=\,2 rotational transitions. Blanked channels correspond to negative features created in the folding of
the frequency switching data. The label U corresponds to features above 4$\sigma$.} \label{spectra}
\end{figure*}

The identification of most of the features from our TMC-1 Q-band line survey
was done using the MADEX code \citep{Cernicharo2012} and the CDMS and
JPL catalogues \citep{Muller2005,Pickett1998}. Nevertheless many lines remain
unidentified. Among these U-lines we found a series of five lines with a
harmonic relation 8:9:10:11:12 between them. This series of lines could be
fitted using a Hamiltonian for a linear molecule obtaining accurate values
for $B$ and $D$ constants;$B$ = 1989.428172 $\pm$ 0.000610 MHz and $D$ =
0.10950 $\pm$ 0.00269 kHz. However, a deeper inspection of the survey around
the mentioned lines revealed the presence of two additional series of lines at
higher and lower frequencies from the first series. The spectral pattern,
taking into account all the lines, is easily recognizable as the typical
$a$-type transition spectrum of a near-prolate molecule, with the sets of rotational
transitions containing
$J$+1$_{0,J+1}$$\leftarrow$$J_{0,J}$, $J$+1$_{1,J+1}$$\leftarrow$$J_{1,J}$
and $J$+1$_{1,J}$$\leftarrow$$J_{1,J-1}$ separated by $B+C$.
All the observed lines, shown in Table \ref{freq_lines} and Fig. \ref{spectra},
were analyzed using an asymmetric rotor Hamiltonian in the FITWAT
code \citep{Cernicharo2018} to derive the rotational and centrifugal distortion
constants shown in Table \ref{rot_const}. With the available data we could not
determine the $A$ rotational constant, which was kept fixed to the ab initio value,
as explained below.

\begin{table}
\tiny
\caption{Observed line parameters for CH$_2$DC$_3$N in TMC-1.}
\label{freq_lines}
\centering
\begin{tabular}{lcccc}
\hline
\hline
{$(J_{K_{\rm a},K_{\rm c}})_{\rm u}$-$(J_{K_{\rm a},K_{\rm c}})_{\rm l}$} & $\nu_{obs}$~$^a$ & $\int T_A^* dv$~$^b$ & $\Delta v$~$^c$ & $T_A^*$ \\
                     &  (MHz)              & (mK\,km\,s$^{-1}$)      & (km\,s$^{-1}$)  & (mK) \\
\hline
$ 8_{1, 8}- 7_{1, 7}$  &  31797.891   & 2.08$\pm$0.12 & 0.84$\pm$0.15& 2.3$\pm$0.4 \\
$ 8_{0, 8}- 7_{0, 7}$  &  31830.626   & 2.76$\pm$0.07 & 0.79$\pm$0.11& 3.3$\pm$0.4 \\
$ 8_{1, 7}- 7_{1, 6}$  &  31862.950   & 1.39$\pm$0.20 & 0.70$\pm$0.12& 1.9$\pm$0.4 \\
$ 9_{1, 9}- 8_{1, 8}$  &  35772.575   & 1.67$\pm$0.14 & 0.64$\pm$0.11& 2.5$\pm$0.4 \\
$ 9_{0, 9}- 8_{0, 8}$  &  35809.389   & 3.11$\pm$0.04 & 0.69$\pm$0.06& 4.3$\pm$0.3 \\
$ 9_{1, 8}- 8_{1, 7}$  &  35845.744   & 1.31$\pm$0.16 & 0.74$\pm$0.14& 3.1$\pm$0.3 \\
$10_{1,10}- 9_{1, 9}$  &  39747.236   & 0.73$\pm$0.30 & 0.45$\pm$0.15& 1.6$\pm$0.4 \\
$10_{0,10}- 9_{0, 9}$  &  39788.132   & 2.66$\pm$0.06 & 0.67$\pm$0.07& 3.8$\pm$0.4 \\
$10_{1, 9}- 9_{1, 8}$  &  39828.533   & 1.07$\pm$0.34 & 0.62$\pm$0.16& 1.6$\pm$0.4 \\
$11_{1,11}-10_{1,10}$  &  43721.867   & 1.45$\pm$0.33 & 0.73$\pm$0.32& 1.9$\pm$0.4 \\
$11_{0,11}-10_{0,10}$  &  43766.845   & 2.11$\pm$0.10 & 0.47$\pm$0.09& 4.2$\pm$0.4 \\
$11_{1,10}-10_{1, 9}$  &  43811.297   & 1.77$\pm$0.24 & 0.98$\pm$0.35& 1.7$\pm$0.4 \\
$12_{1,12}-11_{1,11}$  &  47696.518   & 0.86$\pm$0.40 & 0.82$\pm$0.20& 2.1$\pm$0.5 \\
$12_{0,12}-11_{0,11}$  &  47745.519   & 1.91$\pm$0.10 & 0.46$\pm$0.06& 3.9$\pm$0.5 \\
$12_{1,11}-11_{1,10}$  &  47794.026   & 0.83$\pm$0.46 & 0.45$\pm$0.11& 1.7$\pm$0.5 \\
\hline
\end{tabular}
\tablefoot{
\tablefoottext{a}{Observed frequencies towards TMC-1 for which we adopted a v$_{\rm LSR}$ of 5.83 km s$^{-1}$ \citep{Cernicharo2020a,Cernicharo2020b,Cernicharo2020c}. The
frequency uncertainty 10 kHz.}\tablefoottext{b}{Integrated line intensity in mK\,km\,s$^{-1}$.} \tablefoottext{c}{Line width at half intensity derived by fitting a Gaussian function to the observed line profile (in km\,s$^{-1}$).}
}\\
\end{table}
\normalsize

\begin{table}
\tiny
\caption{Observationally derived and theoretical spectroscopic parameters (in MHz) for CH$_2$DC$_3$N.}
\label{rot_const}
\centering
\begin{tabular}{{lcc}}
\hline
\hline
Constant          &  Space$^a$             & Ab initio$^b$   \\
\hline
$A$               & 120120.70$^c$          &     120120.70    \\
$B$               & 1993.493107(593)       &      1993.53     \\
$C$               & 1985.363112(657)       &      1985.74     \\
$\Delta_{J}$      & 9.107(230) 10$^{-5}$   &       -          \\
$\Delta_{JK}$     & 1.4767(249)10$^{-2}$   &       -          \\
$rms$$^d$         &   13.6                 &       -          \\
$J_{min}/J_{max}$ &   8/12                 &       -          \\
$K_{min}/K_{max}$ &   0/1                  &       -          \\
N$^e$             &   15                   &       -          \\
\hline
\end{tabular}
\tablefoot{
	\tablefoottext{a}{Fit to the lines of CH$_2$DC$_3$N observed in TMC-1.}	\tablefoottext{b}{CCSD/cc-pVTZ level of theory. Scaled values using as reference
    CH$_3$C$_3$N. See text.} \tablefoottext{c}{Fixed to the calculated value.} \tablefoottext{d}{The standard deviation of the fit in kHz.} \tablefoottext{e}{Number of lines included in the fit.}
    }
\end{table}
\normalsize

The identification of the spectral carrier is first based on the following
points. (i) The molecule is a closed-shell species without any appreciable
fine or hyperfine interaction or large amplitude motion. (ii) The
determined values for $B$ and $C$ constants indicate that the molecule
is a very slightly asymmetric rotor, because $B$ and $C$ values are similar. (iii) The ($B$+$C$)/2 value (1989.43 MHz, corresponding to the rotational constant of the close symmetric species) is smaller than that of H$_2$C$_5$ (2295.29 MHz) but larger than that of H$_2$C$_6$ (1344.72 MHz), which indicates that the molecule should contain four C atoms and one atom heavier than C. With these assumptions, we excluded species with four C atoms
and a sulfur atom because they are too heavy. Species containing four C atoms and oxygen, like HC$_4$O \citep{Kohguchi1994} and H$_2$C$_4$O \citep{Brown1979} are rejected as candidates because they are too light ($B$=2279.914 MHz and ($B$+$C$)/2=2153.75 MHz, respectively) and other species derived from them are too heavy or open-shell species. Molecules with four C atoms and nitrogen could be good candidates. HC$_4$N \citep{Tang1999} has a rotational constant $B$=2302.398 MHz and ($B$+$C$)/2 values for the cationic species HC$_4$NH$^+$ and H$_2$C$_4$N$^+$ in their $^1\Sigma$ electronic ground states have been calculated to be 2159.3 MHz and
2194.7 MHz, respectively (CCSD/cc-pVTZ level of theory; \citealt{Cizek1969,Dunning1989}). The next member of this hydrogen addition progression is CH$_3$C$_3$N, whose rotational constant is 2065.74 MHz, very close to our ($B$+$C$)/2 value. This fact prompted us to think that the spectral carrier could be the deuterated isotopologue CH$_2$DC$_3$N , an asymmetric rotor, because the H/D interchange breaks the $C_{3v}$ symmetry of CH$_3$C$_3$N.

We performed geometry optimization calculations for CH$_3$C$_3$N and
CH$_2$DC$_3$N, in order to estimate the isotopic shift on the rotational
constants for the CH$_3$C$_3$N/CH$_2$DC$_3$N system. Using experimental/theoretical
ratios is the most common method to predict the expected experimental rotational
constants for an isotopic species of a given molecule when the rotational
constants for its parent species are known.
Hence, we employed the CCSD/cc-pVTZ level of theory \citep{Cizek1969,Dunning1989}
which reproduces well the $B$ rotational constant for CH$_3$C$_3$N, 2058.0 MHz
$vs.$ 2065.74 MHz. The theoretical values for the rotational constants of
CH$_2$DC$_3$N  were then scaled using the experimental/theoretical ratio
obtained from CH$_3$C$_3$N, and the results are shown in Table \ref{rot_const}.
As it can be seen, the predicted values for CH$_2$DC$_3$N perfectly match
with those derived from our fit, which allows us to conclude that the
spectral carrier of our lines is CH$_2$DC$_3$N. It should be noted that the calculations provide the equilibrium values for the rotational constants ($A_e$, $B_e$ and $C_e$), while the experimental values are the ground state rotational constants ($A_0$, $B_0$ and $C_0$). In spite that the equilibrium rotational constants slightly differ from the ground state constants, we can assume similar discrepancies for CH$_3$C$_3$N and CH$_2$DC$_3$N and, thus, the estimated constants for CH$_2$DC$_3$N are essentially unaffected by this fact.

Methyldiacetylene (CH$_3$C$_4$H) is 7.5 times more abundant than, and lines have similar intensities to, CH$_3$C$_3$N \citep{Marcelino2021,Cernicharo2021c}. Hence, it is straightforward to think that spectral signatures of the deuterated species of CH$_3$C$_4$H could also be detected in our line survey. We followed the same strategy used for CH$_2$DC$_3$N to predict the transition frequencies of CH$_2$DC$_4$H. Laboratory values are available for CH$_3$C$_4$D \citep{Heath1955}. We carried out geometry optimization calculations for CH$_3$C$_4$H.
The rotational constants obtained for CH$_2$DC$_4$H are shown in Table \ref{isotopes}. We found only two lines at the predicted transition frequencies corresponding to the 8$_{0,8}$-7$_{0,7}$ and 9$_{0,9}$-8$_{0,8}$ with intensities $\sim$ 1 mK. Other lines predicted in the frequency range of the line survey are below the present sensitivity. We consider that the deuterated isotopologue of CH$_3$C$_4$H is not detected so far (see below). The deuteration of this species is discussed in the following section.

\begin{table}
\tiny
\caption{Predicted spectroscopic constants$^a$ (in MHz) for isotopic species of CH$_3$C$_3$N and CH$_3$C$_4$H.}
\label{isotopes}
\centering
\begin{tabular}{{lc}}
\hline
\hline
Species          &    $B$              \\
\hline
$^{13}$CH$_3$CCCN  &    2010.51        \\
CH$_3$$^{13}$CCCN  &    2054.75        \\
CH$_3$C$^{13}$CCN  &    2065.71        \\
CH$_3$CC$^{13}$CN  &    2048.72        \\
CH$_3$CCC$^{15}$N  &    2011.56        \\
\hline
$^{13}$CH$_3$CCCCH &    1982.60        \\
CH$_3$$^{13}$CCCCH &    2025.34        \\
CH$_3$C$^{13}$CCCH &    2035.71$^c$    \\
CH$_3$CC$^{13}$CCH &    2018.90        \\
CH$_3$CCC$^{13}$CH &    1980.28        \\
\hline
CH$_2$DCCCCH$^b$   & $A$= 120899.40    \\
                   & $B$=   1965.96    \\
                   & $C$=   1958.41    \\
\hline
\end{tabular}
\tablefoot{
	\tablefoottext{a}{For all the $^{13}$C and $^{15}$N species it can be assumed the $A$ value for the corresponding parent species,
    CH$_3$C$_3$N or CH$_3$C$_4$H, 158099.0 and 159140.0 MHz, respectively.}	\tablefoottext{b}{$A$, $B$ and $C$ constants are provided due to
    the $C_s$ symmetry of this species.} \tablefoottext{c}{For this isotopologue laboratory data are available \citep{Cazzoli2008}.
    The experimental rotational constant $B$ is 2035.67752 MHz.}
    }
\end{table}
\normalsize

Considering the intensity of the CH$_3$C$_3$N and CH$_3$C$_4$H lines in
our line survey, we also expected to observe the $^{13}$C isotopologues as well. The
frequency transitions for these species were predicted using the rotational
constants from Table \ref{isotopes}, which were obtained using the same
procedure employed for the CH$_2$DC$_3$N isotopic species. However, we
could not find spectral signatures for any of these species around the predicted
frequencies.

The column density of CH$_2$DC$_3$N has been derived from a rotational diagram
analysis of the observed intensities. We have assumed a source of uniform brightness
with a radius of 40$''$ \citep{Fosse2001}.
We derive $T_r$=8$\pm$0.5\,K and
N(CH$_2$DC$_3$N)=(8.0$\pm$0.4)$\times$10$^{10}$ cm$^{-2}$.
As shown in Fig. \ref{spectra} the agreement between the synthetic spectrum and the
observations is excellent. The column density is not very sensitive to the adopted
value of the rotational temperature between 6 and 10\,K.
For the normal
isotopologue \citet{Marcelino2021} derived a rotational temperature for the $A$
and $E$ species of 6.7$\pm$0.2\,K and of 8.2$\pm$0.6\,K, respectively.
They derived a total column density for CH$_3$C$_3$N of (1.74$\pm$0.1)$\times$10$^{12}$
cm$^{-2}$. Hence, the CH$_3$C$_3$N/CH$_2$DC$_3$N abundance ratio is 22$\pm$2.

The column density of  CH$_3$C$_4$H has been derived by \citet{Cernicharo2021c}
to be (1.30$\pm$0.04)$\times$10$^{13}$ cm$^{-2}$.
Assuming the same rotational temperature for CH$_2$DC$_4$H than for the main isotopologue
\citep{Cernicharo2021c}, we derive a 3$\sigma$ upper limit to its
column density of 3.7$\times$10$^{11}$ cm$^{-2}$. Therefore, the CH$_3$C$_4$H
over CH$_2$DC$_4$H abundance ratio is $\ge$35 (3$\sigma$). For the deuterated species
CH$_3$C$_4$D, for which laboratory spectroscopy is available \citep{Heath1955},
we derive a 3$\sigma$
upper limit to its column density of 9$\times$10$^{10}$ cm$^{-2}$. Hence,
N(CH$_3$C$_4$H)/N(CH$_3$C$_4$D)$\ge$144.

\begin{table}
\tiny
\caption{Deuteration enhancement in TMC-1 for detected molecules compared to our gas phase chemical model.}
\label{tab_H/D}
\centering
\begin{tabular}{{lccc}}
\hline
\hline
         &   Observations &     Model A  & Model B        \\
Molecule &   TMC-1        &  no          &   full         \\
         &                &  scrambling  &  scrambling    \\
\hline
CH$_3$C$_3$N/CH$_2$DC$_3$N     &       $ 22^a$    &   75.8   &   31.4  \\
CH$_3$CN/CH$_2$DCN             &       11$ ^b$    &   15.0   &   15.0  \\
H$_2$CCN/HDCCN                 &        20$^b$    &   23.6   &   23.6  \\
HC$_3$N/DC$_3$N                &       62 $^c$    &   54.9   &   54.5  \\
HNCCC/DNCCC                    &        43$^c$    &   34.6   &   34.5  \\
HCCNC/DCCNC                    &        30$^c$    &   27.2   &   27.1  \\
HC$_5$N/DC$_5$N$^b$            &        82$^c$    &   23.3   &   23.3  \\
c-C$_3$H$_2$/c-C$_3$HD         &        27$^b$    &   45.5   &   45.4  \\
 C$_4$H/C$_4$D                 &       118$^b$    &   55.5   &   55.3  \\
H$_2$C$_4$/HDC$_4$             &        83$^b$    &   49.5   &   49.4  \\
 CH$_3$CCH/CH$_3$CCD           &        49$^b$    &    257   &    264  \\
CH$_3$CCH/CH$_2$DCCH           &        10$^b$    &     76   &     76  \\
 CH$_3$C$_4$H/CH$_2$DC$_4$H    &   $\ge$35$^d$    &     59   &     20  \\
 CH$_3$C$_4$H/CH$_3$C$_4$D     &  $\ge$144$^d$    &    136   &     55  \\
\hline
\end{tabular}
\tablefoot{
	\tablefoottext{a}{This work.} \tablefoottext{b}{\citet{Cabezas2021}.}\tablefoottext{c}{\citet{Cernicharo2020a}.}\tablefoottext{d}{3$\sigma$ upper limit.}
  }
\end{table}
\normalsize

\section{Chemical Modelling}
We further investigate the chemical processes leading to deuterium insertion
in methylcyanoacetylene and methyldiacetylene,
by extending our previous study on H$_2$CCN and HDCCN \citep{Cabezas2021}.
We only consider gas phase mechanisms which allow  quantitative predictions
based on some experimental measurements and theoretical studies. The chemistry
of the different C$_4$H$_3$N and C$_5$H$_4$ isomers
has been discussed recently in \cite{Cernicharo2021c} and
\cite{Marcelino2021} respectively, in relation with their
detection in TMC-1. These both chemical families are tightly linked
to the chemistry of methylacetylene, CH$_3$CCH and its isomer allene , CH$_2$CCH$_2$.
\begin{eqnarray*}
\rm{CN}  +  CH_3CCH &\rightarrow &  \rm{CH_3C_3N} + H  \\
                &\rightarrow &  \rm{HCN} + CH_2CCH  \\
           &\rightarrow &  \rm{HC_3N} + CH_3  \\
\rm{C_2H}  + CH_3CCH &\rightarrow  &  \rm{CH_3C_4H} + H  \\
           &\rightarrow &  \rm{H_2CCCHCCH} + H
\end{eqnarray*}
whereas
\begin{eqnarray*}
\rm{CN} + CH_2CCH_2  &\rightarrow & \rm{CH_2CCHCN} + H  \\
                                   &  \rightarrow  & \rm{HCCCH_2CN}  + H \\
\rm{C_2H} + CH_2CCH_2  &\rightarrow  & \rm{CH_3C_4H} + H  \\
           &\rightarrow &  \rm{H_2CCCHCCH} + H
\end{eqnarray*}
Considering the deuterated analogs of these reactions introduces diverse question marks, i.e., whether the CN reactions proceed without  changing the methyl radical or lead to some scrambling of the hydrogen atoms in a quasi-stationary intermediate followed by different reaction channels.
The first assumption would lead to the following reactions
\begin{eqnarray}
\rm{CN}  +  CH_2DCCH &\rightarrow &  \rm{CH_2DC_3N} + H  \\
\rm{CN}  +  CH_3CCD&\rightarrow &  \rm{CH_3C_3N} + D
\end{eqnarray}
whereas the second option would introduce an additional reaction channel:
\begin{eqnarray}
\rm{CN}  +  CH_2DCCH &\rightarrow &  \rm{CH_3C_3N} + D              \\
\rm{CN}  +  CH_3CCD & \rightarrow &  \rm{CH_2DC_3N} + H
\end{eqnarray}
The case of the reactions involving C$_2$H (C$_2$D) is even more uncertain as an additional H(D) atom is involved,
which leads to a complementary reaction channel.
\begin{eqnarray}
\rm{C_2D} + CH_3CCH  &\rightarrow & \rm{CH_3C_4H} + D  \\
                                   &  \rightarrow  & \rm{CH_3C_4D}  + H \\
                                   &  \rightarrow  & \rm{CH_2DC_4H}  + H \\
\rm{C_2H} + CH_2DCCH   &\rightarrow  & \rm{CH_2DC_4H} + H  \\
           &\rightarrow &  \rm{CH_3C_4D} + H  \\
           &\rightarrow &  \rm{CH_3C_4H} + D
\end{eqnarray}

Similar questions arise in the deuteration mechanisms involving deuteron transfer  initiated by reactions with abundant deuterated molecular ions such as H$_2$D$^+$ and DCO$^+$.
As an example, the products of the
$ \rm{H_2D^+ + CH_3C_3N }$ reaction could be $\rm{CH_3C_3ND^+} +H_2$ if the reaction proceeds directly
 or also $\rm{CH_2DC_3NH^+} + H_2$ if an intermediate complex is formed.\footnote{The channel CH$_3$C$_3$NH$^+$ + HD is present in both cases as well.} The following step to form deuterated
methylcyanoacetylene entails dissociative recombination of the molecular ion, where an additional question arises on the branching ratios of the reaction $\rm{CH_2DC_3NH^+}$ + e$^-$ $\rightarrow \rm{CH_2DC_3N } + H$ or/and $\rightarrow \rm{CH_3C_3N } + D$.
These few examples show the multiple issues that emerge when analyzing the potential chemical processes at work. We have considered two different approaches: In case A, we assume that the methyl radical and its deuterated form keep their structure (e.g. as in reactions 1,  2, 5, 6, 8) whereas case B involves a scrambling of the H and D atoms followed by the formation of the different products 
(e.g. as in reactions 3, 4, 7, 9, 10). These hypotheses are implemented in a chemical model including 320 species and more than 9000 gas phase reactions built from previous studies
\citep{Cabezas2021,Agundez2021b}. We display in Table \ref{tab_H/D} the corresponding steady state ratios obtained in a model adapted to TMC-1 conditions, n(H$_2$)= 4 $\times 10^4$ cm$^{-3}$, T=10 K, $\zeta$ = 1.3$\times$10$^{-17}$ s$^{-1}$ as in \cite{Cabezas2021}{\footnote{The elemental values, i.e. O/H = 8 10$^{-6}$, C/O = 0.75, N/O= 0.5 correspond to a carbon-rich environment.}}. We first notice the significant sensitivity of the deuterium ratio of
CH$_3$C$_3$N and CH$_3$C$_4$H to the reactivity assumptions. A low deuterium fraction, close to the observed value of CH$_3$C$_3$N, is favored in the full scrambling approximation. The upper limits found for CH$_3$C$_4$H/CH$_2$DC$_4$H and CH$_3$C$_4$H/CH$_3$C$_4$D
are on the other hand better reproduced when reactions 5, 6 and 8 are the only  channels in the C$_2$H (C$_2$D) reactions.
Whereas the other observed deuterium ratios are reasonably reproduced within a factor 2, a significant discrepancy is still obtained for methylacetylene,
CH$_3$CCH, as already noticed in \cite{Cabezas2021,Agundez2021b}.
This feature arises
because the reaction of CH$_3$CCH with H$_3^+$ (and supposedly H$_2$D$^+$) leads to the break up of CH$_3$CCH into $c$-C$_3$H$_3^+$ and $l$-C$_3$H$_3^+$ ($c$-C$_3$H$_2$D$^+$ and $l$-C$_3$H$_2$D$^+$) rather than to C$_3$H$_5$$^+$ or C$_3$H$_4$D$^+$ (CH$_3$CCH$_2$$^+$, CH$_2$DCCH$_2$$^+$, CH$_3$CCHD$^+$), as found in the experimental study of \cite{Milligan2002}. We did not try to include additional deuterium exchange reactions, in the absence of any theoretical or experimental information.

We conclude this section by acknowledging the possible gas phase deuteration mechanisms of
cyanomethylacetylene mediated by deuteron transfer reactions with species such as H$_2$D$^+$, DCO$^+$, among other deuterated cations,
in low temperature conditions but point out the substantial uncertainties involved in the different possible reactions, so
that a detailed comparison between observations and chemical modeling appears elusive.
A theoretical analysis of the intermolecular interaction potentials involved in the
approach of the different reactants would help to validate the various reaction
mechanisms.
\section{Conclusions}
 We have detected, and unambiguously identified, CH$_2$DC$_3$N in TMC-1, a new deuterated compound, thanks to highly sensitive space observations of 15 different transitions and associated theoretical considerations and quantum mechanical calculations.
 Spectroscopic constants are also provided for that compound and the $^{13}$C and $^{15}$N substitutes, which should help to study those species in the laboratory as well. The observed deuterium fractions are further compared to a gas phase model, which, despite significant uncertainties, accounts within a factor of two for the different values, except for CH$_3$CCH. Further experimental or theoretical studies are welcome.

\begin{acknowledgements}
We thank ERC for funding through grant ERC-2013-Syg-610256-NANOCOSMOS.
The Spanish authors thank Ministerio de Ciencia e Innovaci\'on for funding support
through project AYA2016-75066-C2-1-P, PID2019-106235GB-I00 and PID2019-107115GB-C21
/ AEI / 10.13039/501100011033. MA thanks Ministerio de Ciencia e Innovaci\'on for
grant RyC-2014-16277. ER acknowledges the support of the Programme National 'Physique et
Chimie du Milieu Interstellaire' (PCMI) of CNRS/INSU with INC/INP co-funded
by CEA and CNES.
Several kinetic data we used have been taken from the online databases KIDA (\cite{Wakelam2012}, http://kida.obs.u-bordeaux1.fr)
and UMIST2012 (\cite{McElroy2013}, http://udfa.ajmarkwick.net).
\end{acknowledgements}


\begin{thebibliography}{}
\tiny

\bibitem[Ag\'undez et al.(2021a)]{Agundez2021a} Ag\'undez, M., Cabezas, C., Tercero B. et al. 2021a, \aap, 647, L10
\bibitem[Ag\'undez et al.(2021b)]{Agundez2021b} Ag\'undez, M., Roueff, E., Cabezas, C., et al. 2021b, \aap, In press. DOI: https://doi.org/10.1051/0004-6361/202140843.
\bibitem[Broten(1984)]{Broten1984} Broten, N. W., MacLeod, J. M., Avery, L. W., et al. 1984, ApJ, 276, L25
\bibitem[Brown et al.(1979)]{Brown1979}Brown, R. D., Brown, R. F. C., Eastwood, F. W., et al. 1979, J. Am. Chem. Soc. 101, 4705.
\bibitem[Cabezas et al. (2021)]{Cabezas2021} Cabezas, C., Endo, Y., Roueff, E., et al. 2021, \aap, 646, L1
\bibitem[Cazzoli et al. (2008)]{Cazzoli2008}Cazzoli, G., Cludi, L., Contento, M. \& Puzzarini, C. 2008, \jms, 251, 229
\bibitem[Cernicharo(1985)]{Cernicharo1985} Cernicharo, J. 1985, Internal IRAM report (Granada: IRAM)
\bibitem[Cernicharo(2012)]{Cernicharo2012} Cernicharo, J. 2012, in European Conference on Laboratory Astrophysics, eds. C. Stehl\'e, C. Joblin, \& L. d'Hendecourt, EAS Publication Series, 58, 251
\bibitem[Cernicharo et al.(2018)]{Cernicharo2018} Cernicharo, J., Gu\'elin, M., Ag\'undez, M., et al., 2018, A\&A, 618, A4
\bibitem[Cernicharo et al.(2020a)]{Cernicharo2020a} Cernicharo, J., Marcelino, N., Ag\'undez, M., et al. 2020a, A\&A, 642, L17
\bibitem[Cernicharo et al.(2020b)]{Cernicharo2020b} Cernicharo, J., Marcelino, N., Pardo, J. R., et al. 2020b, A\&A, 641, L9
\bibitem[Cernicharo et al.(2020c)]{Cernicharo2020c} Cernicharo, J., Marcelino, N., Ag\'undez, M., et al. 2020c, A\&A, 642, L8
\bibitem[Cernicharo et al.(2021a)]{Cernicharo2021a} Cernicharo, J., Cabezas, C., Ag\'undez, M., et al., 2021a, \aap, 648, L3 
\bibitem[Cernicharo et al.(2021b)]{Cernicharo2021b} Cernicharo, J., Ag\'undez, M., Cabezas, C., et al. 2021b, \aap, 647, L2 
\bibitem[Cernicharo et al.(2021c)]{Cernicharo2021c} Cernicharo, J., Cabezas, C., Ag\'undez, M., et al., 2021c, \aap, 647, L3 
\bibitem[Cernicharo et al.(2021d)]{Cernicharo2021d} Cernicharo, J., Ag\'undez, M., Cabezas, C., et al. 2021d, \aap, 649, L15 
\bibitem[Cernicharo et al.(2021e)]{Cernicharo2021e} Cernicharo, J., Cabezas, C., Endo, Y., et al. 2021e, \aap, 646, L3 
\bibitem[Cernicharo et al.(2021f)]{Cernicharo2021f} Cernicharo, J., Cabezas, C., Bailleux, S., et al. 2021f, \aap, 646, L7 
\bibitem[C\'i\v{z}ek(1969)]{Cizek1969} C\'i\v{z}ek, J., in "Advances in Chemical Physics" (P. C. Hariharan, Ed.), Vol.14, 35, Wiley Interscience, New York, 1969
\bibitem[Dunning(1989)]{Dunning1989}Dunning, T.~H., 1989, J. Chem. Phys. 90, 1007
\bibitem[Foss\'e et al.(2001)]{Fosse2001} Foss\'e, D., Cernicharo, J., Gerin, M., Cox, P. 2001, \apj, 552, 168
\bibitem[Kohguchi et al.(1994)]{Kohguchi1994} Kohguchi, H., Ohshima, Y., \& Endo, Y. 1994, J. Chem. Phys. 101, 6463.
\bibitem[Heath et al.(1955)]{Heath1955} Heath, G.A., Thomas, L.F., Sherrard, E.I., \& Sheridan, J. 1955, Faraday Soc. Disc., 19, 38
\bibitem[Hickson, Wakelam and Loison (2016)]{Hickson2016} Hickson, K.B., Wakelam, V., Loison, J.C., 2016, Molec. Astrophys. 3, 1
\bibitem[Linsky(2003)]{Linsky2003} Linsky, J. L. 2003, \ssr, 106, 49
\bibitem[Marcelino et al.(2005)]{Marcelino2005} Marcelino, N., Cernicharo, J., Roueff, E., et al. 2005, \apj, 620, 308
\bibitem[Marcelino et al.(2021)]{Marcelino2021} Marcelino, N., Tercero, B., N., Ag\'undez, M., \& Cernicharo, J., et al. 2021, \aap, 646, L9
\bibitem[McElroy et al.(2013)]{McElroy2013} McElroy, D., Walsh, C., Markwick, A. J., et al. 2013, \aap, 550, A36
\bibitem[Milligan et al.(2002)]{Milligan2002} Milligan, D.B., Wilson, P.F., Freeman, C.G., et al. 2002, J. Phys. Chem. A, 106, 9745
\bibitem[M\"uller et al.(2005)]{Muller2005} M\"uller, H. S. P., Schl\"oder, F., Stutzki, J., \& Winnewisser, G. 2005, \jmst, 742, 215
\bibitem[Parise et al.(2006)]{Parise2006} Parise, B., Ceccarelli, C., Tielens, A. G. G. M., et al. 2006, \aap, 453, 949
\bibitem[Tang et al.(1999)]{Tang1999}Tang, J., Sumiyoshi, Y., \& Endo, Y. 1999, Chem. Phys. Lett., 315, 69
\bibitem[Tercero et al. (2021)]{Tercero2021} Tercero, F., L\'opez-P\'erez, J. A., Gallego, et al., 2021, \aap, 645, A37
\bibitem[Pardo et al.(2001)]{Pardo2001} Pardo, J.~R., Cernicharo, J., Serabyn, E. 2001, IEEE Trans. Antennas and Propagation, 49, 12
\bibitem[Pickett et al.(1998)]{Pickett1998} Pickett, H. M., Poynter, R. L., Cohen, E. A., et al. 1998, J. Quant. Spectr. Rad. Transfer, 60, 883
\bibitem[Wakelam et al.(2012)]{Wakelam2012} Wakelam, V., Loison, J.-C., Herbst, E., et al. 2012, \apjs, 199, 21

\end{thebibliography}
\end{document}